# Gate-tunable negative refraction of mid-infrared polaritons

Hai Hu,[1, 2†*] Na Chen,[1,2†] Hanchao Teng,[1, 2†] Renwen Yu,[3] Mengfei Xue,[6] Ke Chen, [1, 2] Yuchuan Xiao,[1, 2] Yunpeng Qu,[1, 2] Debo Hu,[1, 2] Jianing Chen,[6] Zhipei Sun,[7] Peining Li,[8] F. Javier García de Abajo,[4, 5*] Qing Dai[1, 2*]

1 CAS Key Laboratory of Nanophotonic Materials and Devices, CAS Key Laboratory of Standardization and Measurement for Nanotechnology, CAS Center for Excellence in Nanoscience, National Center for Nanoscience and Technology, Beijing 100190, P. R. China.

2 University of Chinese Academy of Sciences, Beijing 100049, P. R. China.

3 Department of Electrical Engineering, Ginzton Laboratory, Stanford University, Stanford, CA, USA.

4 ICFO-Institut de Ciencies Fotoniques, The Barcelona Institute of Science and Technology, 08860 Castelldefels (Barcelona), Spain.

5 ICREA-Institució Catalana de Recerca i Estudis Avançats, Passeig Lluís Companys 23, 08010 Barcelona, Spain.

6 The Institute of Physics, Chinese Academy of Sciences, P.O. Box 603, Beijing, China.

7 Department of Electronics and Nanoengineering, Aalto University, FI-02150 Espoo, Finland.

8 Wuhan National Laboratory for Optoelectronics and School of Optical and Electronic Information, Huazhong University of Science and Technology, Wuhan, P. R. China.

*e-mail: daiq@nanoctr.cn, javier.garciadeabajo@nanophotonics.es, huh@nanoctr.cn

† These authors contributed equally

## Keywords

## Abstract

**Negative refraction provides an attractive platform to manipulate mid-infrared and terahertz radiation for molecular sensing and thermal radiation applications. However, its implementation based on available metamaterials and plasmonic media presents challenges associated with optical losses, limited spatial confinement, and lack of active tunability in this spectral range. Here, we demonstrate gate-tunable negative refraction at mid-infrared frequencies using hybrid topological polaritons in van der Waals heterostructures with high spatial confinement. We experimentally visualize wide-angle negatively-refracted surface polaritons on α-MoO3 films partially decorated with graphene, undergoing planar nanoscale focusing down to 1.6% of the free-space wavelength. Our atomically thick heterostructures outperform conventional bulk materials by avoiding scattering losses at the refracting interface while enabling active tunability through electrical gating. We propose polaritonic negative refraction as a promising platform for infrared applications such as electrically tunable super-resolution imaging, nanoscale thermal manipulation, and molecular sensing.**

## Introduction

Negative refraction has been extensively investigated in optics(*1-3*), nanoelectronics(*4*), acoustics(*5*), and magnetism(*6*) as a counter-intuitive physical phenomenon—bending of electromagnetic fields in the 'wrong' direction—that holds strong potential for applications such as subwavelength imaging and cloaking(*7*). The last two decades have witnessed substantial progress in the study of negative refraction, typically implemented using metallic metamaterials(*8, 9*), dielectric photonic crystals(*10, 11*), and hyperbolic metamaterials(*12, 13*) that are composed of periodic arrays of subwavelength unit cells and possess extraordinary optical properties beyond those existing in conventional materials. The metamaterial concept invoked in these structures (*i.e.*, their behavior as homogeneous media for small unit cells) limits their ability to strongly confine light. As an alternative, metal plasmons have also been demonstrated to undergo negative refraction in the ultraviolet(*14*), visible(*15*), and near-infrared(*16*) spectral regions. This approach is limited by ohmic losses at visible and higher frequencies as well as poor spatial confinement of plasmons in the infrared range. Deep-subwavelength negative refraction at mid-infrared and terahertz frequencies has therefore remained as a challenge, despite the interest arising from its potential in the manipulation of signals associated with molecular vibrations and thermal emission in such spectral domains.

The emergence of van der Waals (vdW) two-dimensional materials has introduced a new degree of freedom in controlling light at the nanoscale over a wide spectral range by leveraging the strong optical confinement associated with their polaritonic modes(*17-21*). These materials exhibit a combination of tunability, low losses, and ultrahigh confinement that makes them appealing for the design of nanophotonic devices. Understandably, recent theoretical studies have proposed the use of vdW polaritons to achieve deep sub-wavelength mid-infrared negative refraction, for example, in periodic arrays of graphene(*22*), or using subtly planar heterostructures formed by graphene and hexagonal boron nitride (*h*-BN)(*23*). Nevertheless, the extreme spatial confinement of polaritons in these structures makes it difficult to tailor their dispersion relations and achieve negative refraction. In addition, reflection and scattering losses that are inherent to such structures also complicate the realization of this theoretical concept.

Here, we demonstrate deep-subwavelength mid-infrared negative refraction by constructing a vdW heterostructure consisting of an α-MoO$_3$ film partially covered by monolayer graphene. This heterostructure provides hybrid topological polaritons with tunable contours by varying the graphene doping level, such that negative refraction takes place when combining hyperbolic and elliptic dispersion profiles on either side of the in-plane interface. In addition, optical losses are reduced during refraction due to the unique single-atomic-layer nature of graphene and the strong modal-profile matching of the involved polaritons. By adjusting the position of a launching metal antenna used as a source of polaritons, we demonstrate reversible negative refraction over a wide range of incidence angles, allowing us to focus incident polaritons with either concave or convex wavefronts, without suffering undesired diffraction effects.

## Results and discussion

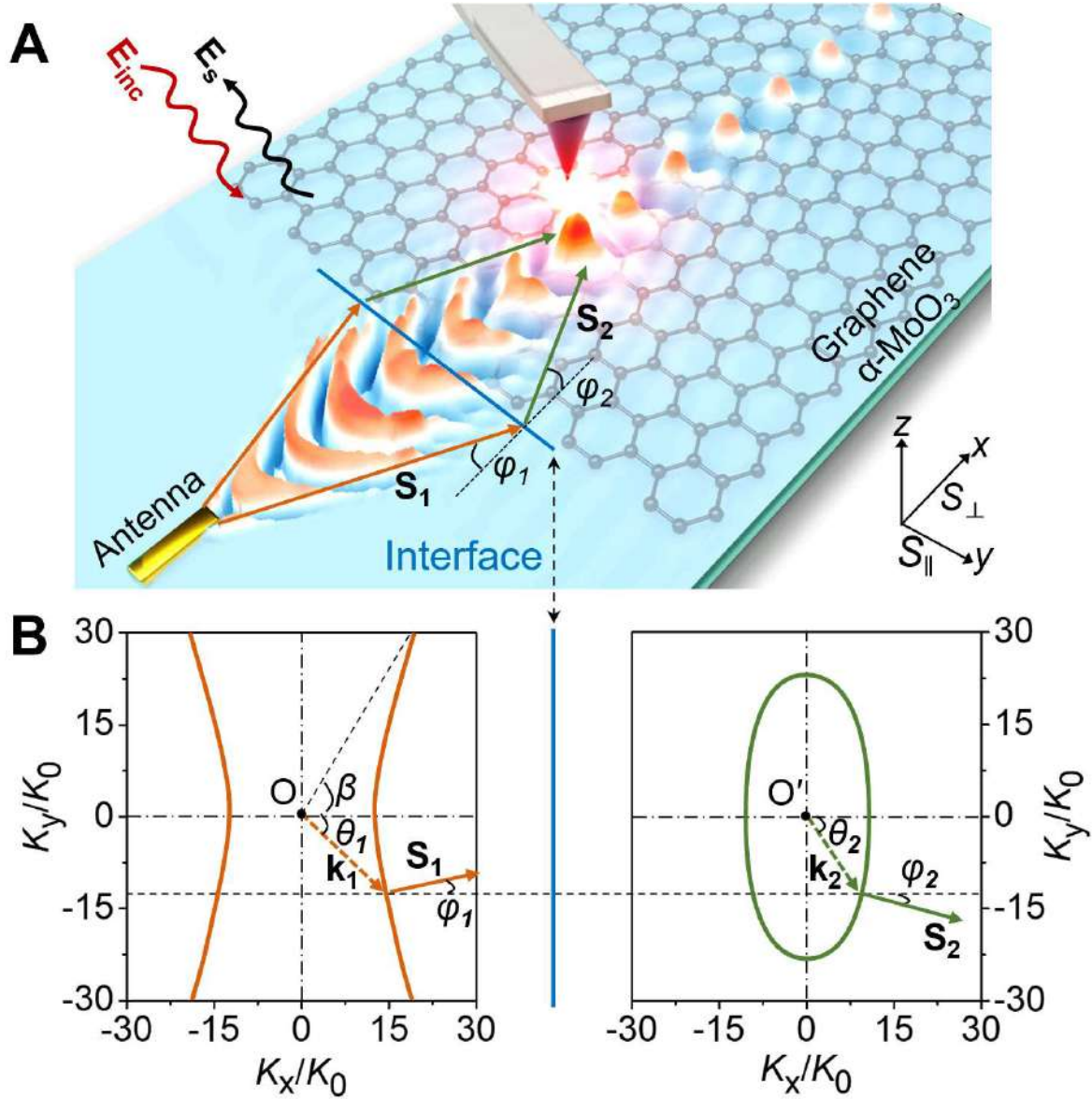

**Figure 1. Negative refraction of polaritons in two dimensions.** (**A**) Schematics of the device. Gold antennas are patterned on the sample to launch polaritons. Features associated with negative refraction are revealed by a scattering-type scanning near-field optical microscope (s-SNOM) equipped with a tunable quantum cascade laser(*24, 25*). (**B**) Isofrequency contours (IFCs) of topological polaritons in hyperbolic (α-MoO₃) and elliptic (graphene/α-MoO₃) media. Negative refraction of a wave propagating in the former (solid orange arrow) takes place at the interface between the two media due to conservation of parallel wave vector (indicated by dashed arrows), which results in a transmitted wave propagating on the other side of the interface normal (solid green arrow). We define incidence and refraction angles $\theta_1$ and $\theta_2$ of the polaritons with wave vectors $\mathbf{k}_1$ and $\mathbf{k}_2$, as well as incidence and refraction angles $\varphi_1$ and $\varphi_2$ of the corresponding polariton Poynting vectors $\mathbf{S}_1$ and $\mathbf{S}_2$.

Phonon polaritons (PhPs) in α-MoO₃ exhibit hyperbolic in-plane dispersion in the Reststrahlen band II from 816 cm$^{-1}$ to 972 cm$^{-1}$, where the permittivity components along the [100], [001], and [010] crystal directions satisfy $\varepsilon_x < 0$, $\varepsilon_y > 0$, and $\varepsilon_z > 0$, respectively([26-29]). In contrast, graphene supports highly confined isotropic plasmons in this spectral region([30, 31]), such that anisotropic hyperbolic PhPs in the graphene/α-MoO₃ heterostructure couple to graphene plasmons and undergo an optical topological transition that allows us to flexibly engineer the polariton dispersion and IFCs([32-35]).

By conserving the polariton wave vector along the direction of the graphene edge, negative refraction is shown to take place when polaritons traverse the in-plane interface between the two regions (Figure 1A). The Poynting vector **S** (directed along the energy flow, which is normal to the IFC) and wave vector **k** (normal to the wavefront) are not collinear for non-circular IFCs (Supplementary Figure 1 and Note 1), such as those of hyperbolic polaritons in α-MoO₃ (Figure 1B)([36]). Since the boundary conditions at the interface (between α-MoO₃ with and without covering graphene) only require conservation of the tangential wave vector component $k_{/\!/} = k \sin\theta$, where $\theta$ is the angle between the wave vector and the interface normal, the refracted wave can exhibit standard (positive) refraction for **k**, but negative refraction for **S** (with $S_{/\!/} = S \sin\varphi$, where $\varphi$ is the angle between the Poynting vector and the interface normal).

Negative refraction transforms the linear interface into a lens capable of focusing polaritons at a position determined from the relation between the incidence and refraction angles, which can in turn be obtained by inspecting the distribution of the *z-z* component of the Dyadic Green function in real space and then comparing the Poynting vector **S** to the IFCs (see details in Supplementary Figure 2 and Note 2). Notably, because the IFCs possess inversion symmetry with respect to the graphene edge direction, negative refraction occurs reversibly when crossing the interface in both directions (*i.e.*, from or to bare α-MoO₃).

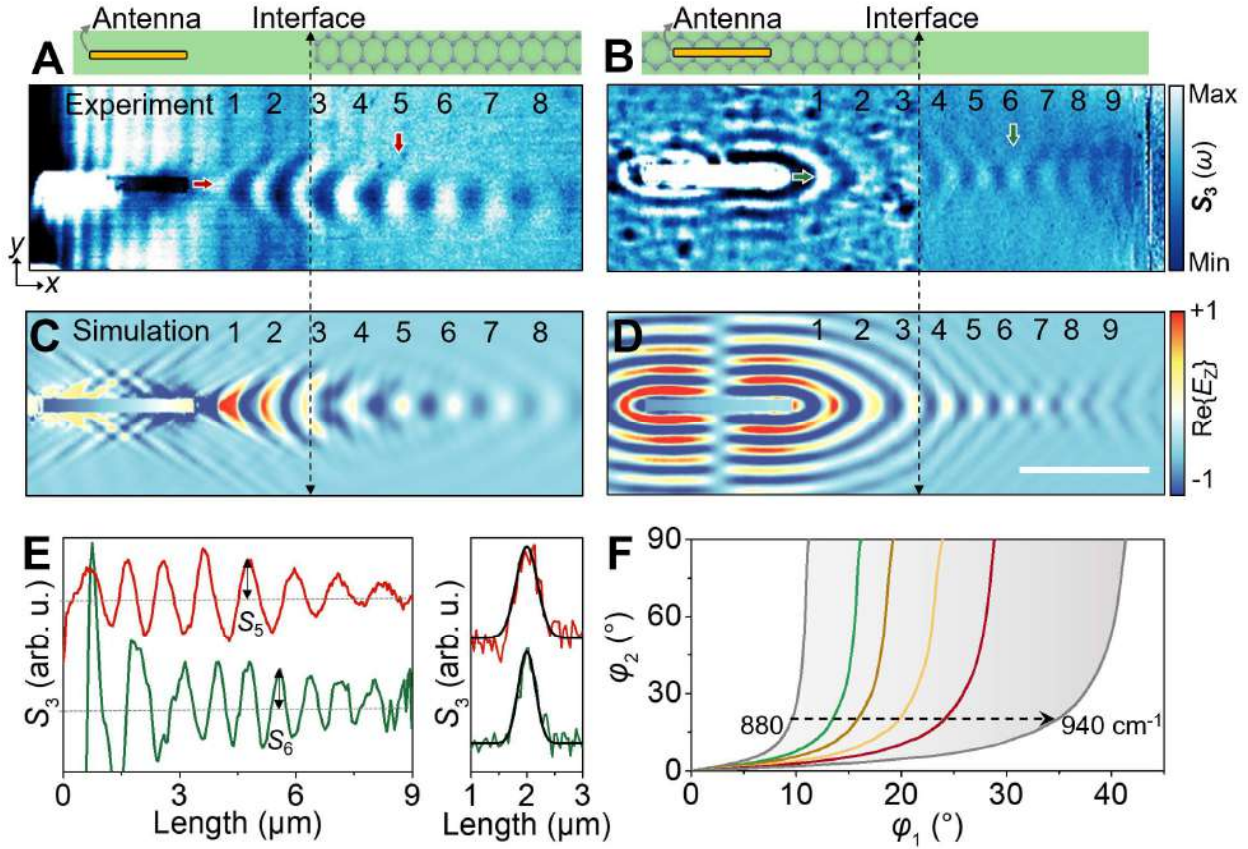

**Figure 2. Direct observation of nanoscale negative refraction.** (**A, C**) Experimental (A) and simulated (C) near-field images illustrating negative refraction from a hyperbolic wave in α-MoO₃ (launched from a gold antenna, see upper sketch) to an elliptic wave in graphene/α-MoO₃. (**B, D**) Negative refraction from elliptic to hyperbolic waves with the antenna now placed on the graphene/α-MoO₃ side. The scale bar indicates 3 µm. (**E**) Near-field profiles along the *x* (left) and y (right) directions (see arrows in panels (A) and (B)). Black curves in the latter are Gaussian fits of the transverse focal profile. All experimental results are measured from an *in-situ* sample. The α-MoO₃ thickness is *d* = 242 nm. The graphene is mechanically exfoliated and statically doped to $E_F$ = 0.5 eV by coverage with monolayer α-RuCl₃. The illumination frequency is $\omega_0$ = 893 cm⁻¹ (A) and 900 cm⁻¹ (B), respectively. (**F**) Refraction (incidence) angle $\varphi_2$ in graphene/α-MoO₃ as a function of incidence (refraction) angle $\varphi_1$ in α-MoO₃ for various illumination frequencies from 880 to 940 cm⁻¹(gray shaded area surrounded by dashed curves). Color curves represent specific frequencies measured in our experiments.

For an experimental demonstration of negative refraction, we fabricate a gold antenna on one side of the in-plane interface (Supplementary Figure 3), serving as a source of polaritons. When hyperbolic PhPs are launched on the bare α-MoO₃ side and propagate towards the α-MoO₃ region covered by graphene (where elliptic polaritons arise from hybridization with graphene plasmons), negative refraction occurs and the transmitted concave wavefront is observed to shrink sharply (Figures 2A, C). The focal point is situated at the fringe with the smallest FWHM (the fifth fringe), yielding foci of FWHM as small as 373 nm (*i.e.*, 1/30 of the free-space light wavelength) (Figure 2E). After passing the focal spot, the wavefront begins to spread slightly in a diffractive fashion (sixth and higher order of fringes).

To study the predicted reversibility of negative refraction *in situ*, we have developed an AFM probe transfer method to remove the antenna and subsequently fabricate another one at a different position on the same sample (Supplementary Figure 4). Following this approach, we also demonstrate negative refraction when reversing the propagation direction by placing the launching antenna on the graphene/α-MoO$_3$ side (Figures 2B, D), so that the diverging elliptic polaritons in such medium are transmitted into converging hyperbolic PhPs in bare α-MoO$_3$, and a tightly squeezed focal spot of FWHM = 303 nm is produced (Figure 2E). After passing the focal point, the wavefront begins to spread slightly, but now with a hyperbolic wavefront. For a given set of polariton IFCs, the focusing effect can be modulated by moving the launching source, which produces a phase shift of the entire propagation wave (Supplementary Figure 5).

The focusing concentrates the energy carried by polaritons to enhance the field intensity. Indeed, the square of the ratio of the electrical field at the focal spot to that without focusing yields a 10-fold increase in intensity (Figure 2E and Supplementary Figure 6). We note that losses associated with refraction at the interface, which involve modal-profile mismatch between PhPs in α-MoO$_3$ and polaritons in graphene/α-MoO$_3$, as well as additional losses at the graphene edge are relatively low (see details in Supplementary Figures 7, 8). Our experimental measurements agree well with simulated near-field distributions of Re{$E_z$} (Figures 2A-D) and the extracted simulation near-field profile also quantitatively matches the experimental results (Supplementary Figure 6).

The reported in-plane negative refraction effect takes place over a wide range of incidence angles (Figure 2F) as well as spectral range, where the opening angle $\beta$ of PhPs in α-MoO$_3$ increases with the illumination frequency and the focal length varies in opposite ways with frequency for the two refraction scenarios (Supplementary Figures 9, 10). We plot the relationship between incidence and refraction angles at different illumination frequencies ranging from 880 to 940 cm$^{-1}$. For incidence or refraction in bare α-MoO$_3$, the incidence angle $\varphi_1$ or the refraction angle $\varphi_2$, respectively, are limited by the opening angle $\beta$ because PhP propagation is prohibited beyond the hyperbolic region of the IFCs, whereas negative refraction can be generated for a wide range. Notably, in the Reststrahlen band III (higher light frequency from 972 to 1005 $cm^{-1}$), a lateral heterostructure can also achieve nanoscale negative refraction arising from the negative group velocity associated with the reversed dispersion of PhPs in α-MoO₃ (Supplementary Figure 11) (*23*).

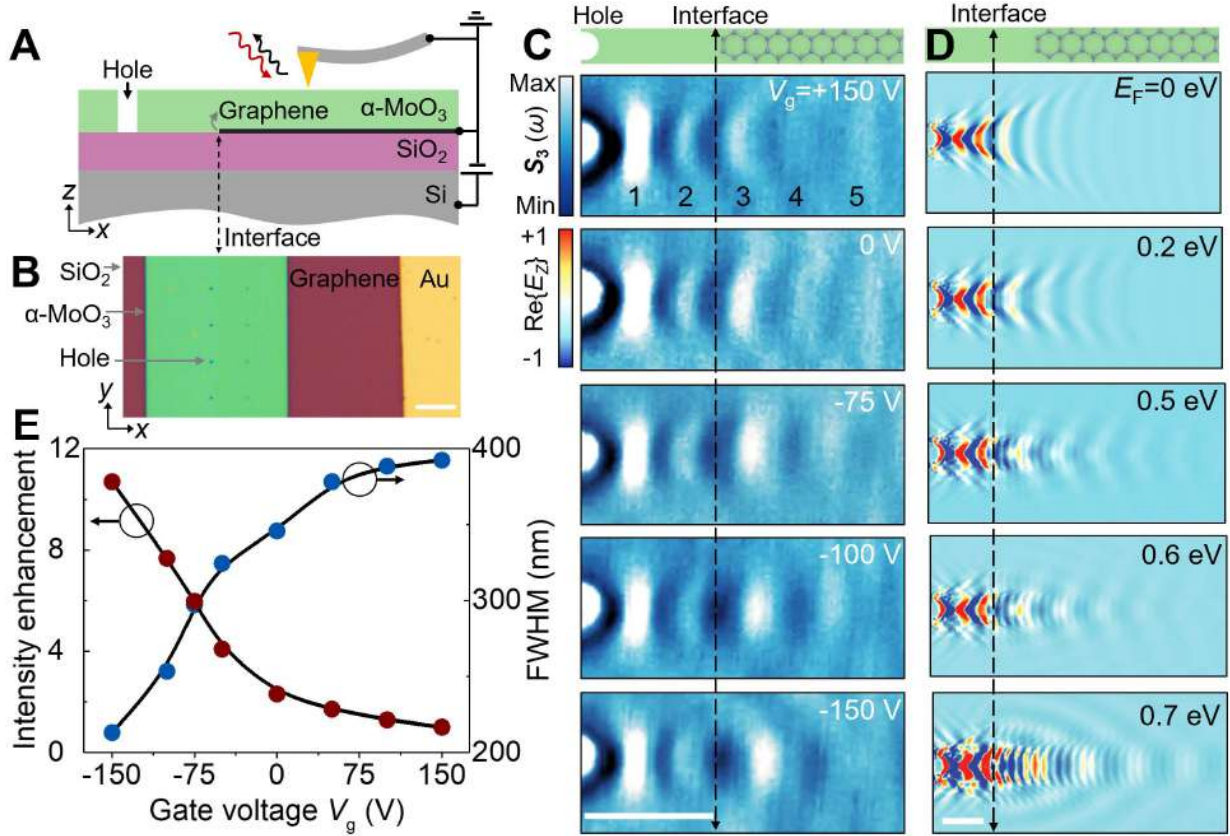

**Figure 3. Gate-tunable nanofocusing of negative refraction. (A)** Scheme of the device, in which polaritons are launched by the tip and reflected by a hole in α-MoO₃. (**B**) Optical image of a gate-tunable device consisting of (from top to bottom) an α-MoO₃ film, bilayer graphene, and SiO₂ substrate. The grey horizontal arrow indicates the position of fabricated circular holes with a diameter of 400 nm. (**C**) Experimentally measured near-field images of gate-tunable negative refraction from hyperbolic α-MoO₃ to elliptic graphene/α-MoO₃ with bias voltages varying from +150 V to -150 V. The vertical black dashed line represents the graphene edge. (**D**) Numerically simulated negative refraction with various Fermi energies of graphene from $E_F$=0 to 0.7 eV. (**E**) Intensity enhancement and FWHM of the focal spots for various gate voltages, taken from the experimental measurements in panel (C). The α-MoO₃ thickness is $d$ = 60 nm and the illumination wavelength is fixed at $\lambda_0$ = 11.20 μm (frequency of 893 $cm^{-1}$). The scale bars indicate 10 μm and 1 μm in panels (B) and (C, D), respectively.

Active control of negative refraction can be gained via electrostatic gating by varying the graphene chemical potential. We have demonstrated this possibility by preparing samples with a $SiO_2$/doped-Si substrate instead of Au, such that we can apply a perpendicular electric field to graphene/α-$MoO_3$ via a $SiO_2$ dielectric layer coating a Si back-gate (Figure 3A). To obtain a clean interface, we have directly transferred an α-$MoO_3$ film with through holes on the exfoliated graphene with pre-fabricated electrodes (Figure 3B), where polaritons are excited by the tip and subsequently reflected by the hole edge.

The effect of changing the gating voltage $V_g$ on the polaritons is shown in Figure 3C and Supplementary Figure 12, where the Fermi energy is tuned over a wide range from $E_F$ = 0 to ~0.7 eV. By decreasing $V_g$, the open-angle of the hyperbolic wavefront continues to shrink towards the canalization point for the graphene/α-$MoO_3$ region (Supplementary Figures 12, 13 and Figure 3C), whereas the hyperbolic polaritons in bare α-$MoO_3$ show no response to the gate voltage (Supplementary Figure 13A). This attributes to a gradual switch on of negative refraction that originates in the topological transition of hybrid polaritons as holes are increasingly injected into graphene by lowering $V_g$. Our rigorous simulations further reproduce the continuous switching of negative refraction with varying doping level of graphene (Figure 3D and Supplementary Figure 13B). This gate-tunable negative refraction provides a unique capability to actively control the wavefront of polaritons *in situ* and change the focusing position and the associated optical fields at the nanoscale (Figures 3C, D). As the gate voltage changes from +150 V to -150 V, the FWHM transverse size of the focal spot decreases, and the focal field intensity increases by more than one order of magnitude (Figure 3E and Supplementary Figure 14). Supplementary Figure 15 shows our gate-tunable negative refraction of another sample with monolayer graphene and thinner α-$MoO_3$.

Negative refraction achieved by engaging topological polaritons with different IFC shapes is distinct from that attained by using bulk materials with a negative index of refraction(*8*). We are essentially relying on a homogeneous film (α-$MoO_3$) in which the addition of an atomically thin layer (graphene) radically changes the surface-mode characteristics to enable negative refraction. By changing the Fermi energy of graphene (and, correspondingly, the composition of the topological hybrid polaritons), we can actively and continuously transition from positive to negative refraction, whereas control of bulk materials remains as a challenge. However, to achieve a high degree of negative refraction, the required doping level is relatively high (~0.8 eV) and difficult to reach using a typical $SiO_2$/Si back-gate. Therefore, we use bilayer graphene and adopt thinner α-$MoO_3$ films in Figure 3 to decrease the demand for gate voltages and graphene doping level, although this results in a shorter propagation distance than in the thicker sample in Figure 2. To balance propagation distance and tunability range, more efficient double (top-and-back) gates[37] are promising for future investigations, possibly combined with ion-gel spacers.

## Conclusion

We have experimentally demonstrated mid-infrared negative refraction by designing a vdW heterostructure consisting of an extended α-$MoO_3$ film that is half-covered by monolayer graphene. The observed nanoscale negative refraction and focusing, which relies on topological polaritons with tunable dispersion curves, is revealed through real-space mapping based on infrared nanoscopy. We leverage negative refraction to demonstrate nanoscale optical focusing, which, thanks to the high spatial confinement of polaritons and the atomic thickness of the employed vdW structures, results in deep-subwavelength focal spots with highly squeezed sizes of less than 60 times that of the corresponding illumination wavelength, as well as a 10-fold intensity enhancement and a ~90% transmission of negatively refracted energy (Supplementary Figure 8). Importantly, we show that negative refraction can be actively tuned by an electrostatic gate, resulting in the unique ability to control the wavefront of polaritons *in situ* and change focal spots and their associated nanoscale optical fields.

Considering the vast range of newly available two-dimensional polaritonic materials, we anticipate negative refraction of polaritons in other vdW heterostructures involving, for example, α-$V_2O_5$, black phosphorus, and nanostructured metasurfaces (*e.g.*, based on isotope h-BN). The broad suite of existing materials could lead to polaritonic negative refraction covering the entire mid-infrared and terahertz region. The combined advantage of strong polariton-field confinement, flexible control over anisotropic polariton propagation and focusing, and tunability by material stacking as well as electric gating opens exciting avenues for negative refraction in optical and thermal applications.

## Methods

**Nanofabrication of the devices.**

The α-RuCl₃/graphene/α-MoO₃ heterostructures were exfoliated and stacked using a vdW assembly technique (Supplementary Figure 3). Initially, monolayer α-RuCl₃, monolayer graphene, and α-MoO₃ films (~242 nm) were taken from natural crystals and exfoliated on $SiO_2$ (300nm)/Si and Au (60nm)/Si substrates, respectively. Then, α-RuCl₃ and graphene were picked up from a $SiO_2$/Si substrate by a deterministic dry-transfer process aided by a glass /polydimethylsiloxane (PDMS)/polycarbonate (PC) stamp at 100 ℃. After that, the stamp with α-RuCl₃/graphene was aligned with α-MoO₃ and the temperature was raised to 180 ℃ to release the heterostructure with PC and form α-RuCl₃/graphene/α-MoO₃ heterostructures. The samples were subsequently immersed in chloroform for 10 min and rinsed in IPA for 3 min to eliminate any residual PC. We note that monolayer α-RuCl₃ has been reported to produce high and homogeneous doping in monolayer graphene with mobilities ~4900 $cm^2$/ (V s) at a large hole density of $3 \times 10^{13}$ $cm^{-2}$, while having minimal optical absorption in the mid-infrared range([38]).

The Au antenna arrays (3 μm × 250 nm × 50 nm) were patterned on the sample using 100 kV electron-beam lithography (EBL) (Vistec 5000+ES, Germany) on PMMA950K resist (thickness of ~350 nm). For fabrication of Au antennas, a layer of 50 nm thickness of Au was deposited by electron-beam evaporation in a vacuum chamber under a pressure of $5 \times 10^{-6}$ Torr. We deliberately did not prepare an adhesion layer (usually 5 nm thickness of Ti or Cr), so that the gold antennas on the sample could be moved away by an AFM tip (Supplementary Figure 4). Electron-beam evaporation was also used to deposit a 60 nm-thick gold film onto a low-doped Si substrate. To remove any residual organic materials, the samples were immersed in a hot acetone bath at 80 °C for 25 min and subject to a gentle rinse of IPA for 3 min, followed by nitrogen gas drying and thermal baking.

**Near-field optical microscopy measurements.**

For near-field measurements, a scattering SNOM setup (Neaspec GmbH) equipped with a tunable quantum cascade laser (890 to 2000 $cm^{-1}$) was utilized. A Pt-coated AFM tip with a radius of ~25 nm (NanoWorld) was employed, with a tapping frequency and amplitude of ~270 kHz and ~30-50 nm, respectively. A *p*-polarized mid-infrared beam with a lateral spot size of 25 μm was aimed at the AFM tip, illuminating the antennas and a large area of the graphene/α-MoO$_3$ samples. To effectively reduce background noise, a third-order demodulated-harmonic analysis of the near-field amplitude images was applied.

**Calculation of dispersion and IFCs of hybrid plasmon-phonon polaritons.**

The dispersion relation of hybrid polaritons was rigorously obtained from waveguide theory applied to graphene/α-MoO$_3$ heterostructures, with the structures modeled as a 2D waveguide consisting of four stacked layers (see more details in Supplementary Figure 1 and Note 1). In this analysis, air and the gold substrate were modeled in terms of their isotropic dielectric tensors, while α-MoO$_3$ was described by anisotropic tensors.

**Electromagnetic simulations.**

A Finite Element Method software (COMSOL Multiphysics 5.5) was used to simulate the electromagnetic field. The model was constructed by following the geometrical specifications in the experimental samples, in which graphene was described as a 0.33-nm-thick transition interface, the metallic antenna was placed on the graphene/α-MoO$_3$ sample, and the substrate was a 60-nm-thick gold film. Perfectly matched layers were set up at all boundaries around the model to reduce boundary reflections. The incident light was set as a plane wave polarized along the long axis of the antenna with an angle of 45º to the surface. To simulate the tip-launched polaritons, we introduced a dipole located 100 nm above the surface and polarized perpendicular to the surface. The electric field distribution Re{$E_z$} was calculated on a plane situated 20 nm above the uppermost surface of the sample. To simplify the simulation, we ignored the single layer of α-RuCl$_3$. This approximation should have a negligible effect on the simulation outcomes, as a single layer of α-RuCl$_3$ would only introduce a small dielectric loss (Supplementary Figure 16).

# Acknowledgments

This work was supported by the National Key Research and Development Program of China (grant no. 2021YFA1201500, to Q.D.; 2020YFB2205701, to H.H.), the National Natural Science Foundation of China (grant nos. 51902065, 52172139 to H.H.; 51925203,

U2032206, 52072083 and 51972072, to Q.D.), Beijing Municipal Natural Science Foundation (grant no. 2202062, to H.H.) and the Strategic Priority Research Program of Chinese Academy of Sciences (grant nos. XDB30020100 and XDB30000000, to Q.D.). F.J.G.A. acknowledges the ERC (Advanced grant no. 789104-eNANO), the Spanish MICINN (PID2020-112625GB-I00 and SEV2015-0522), and the CAS President's International Fellowship Initiative for 2021. Z.S. acknowledges the Academy of Finland (grant nos. 314810, 333982, 336144, and 336818), The Business Finland (ALDEL), the Academy of Finland Flagship Programme (320167, PREIN), the European Union's Horizon 2020 research and innovation program (820423, S2QUIP and 965124, FEMTOCHIP), the EU H2020-MSCA-RISE-872049 (IPN-Bio) and the ERC (834742). P.L. acknowledges the National Natural Science Foundation of China (grant no. 62075070).

## Data availability

The data that support the findings of this study are available from the corresponding author upon reasonable request.

## Author contributions

Q.D., F.J.G.A., and H.H. conceived the idea. Q.D. and F.J.G.A. supervised the project. H.H. and N.C. led the experiments, prepared the samples, and performed the near-field measurements. H.T., R.Y., and F.J.G.A. developed the theory and performed the simulation. H.H., N.C., and H.T. analyzed the data, and all authors discussed the results. H.H. and N.C. co-wrote the manuscript, with input and comments from all authors.

## Competing interests

The authors declare no competing interests.

## Reference

1. J. B. Pendary, *Phys. Rev. Lett.* **85**, 3966 (2000).
2. R. A. Shelby, D. R. Smith, S. Schultz, *Science* **292**, 77-79 (2001).
3. J. Yao *et al.*, *Science* **321**, 930 (2008).
4. G.-H. Lee, G.-H. Park, H.-J. Lee, *Nat. Phys.* **11**, 925-929 (2015).
5. H. He *et al.*, *Nature* **560**, 61-64 (2018).
6. A. Pimenov, A. Loidl, P. Przyslupski, B. Dabrowski, *Phys. Rev. Lett.* **95**, 247009 (2005).
7. W. Cai, U. K. Chettiar, A. V. Kildishev, V. M. Shalaev, *Nat. Photon.* **1**, 224-227 (2007).
8. D. R. Smith, J. B. Pendry, M. C. Wiltshire, *Science* **305**, 788-792 (2004).
9. K. L. Tsakmakidis, A. D. Boardman, O. Hess, *Nature* **450**, 397-401 (2007).
10. E. Cubukcu *et al.*, *Nature* **423**, 604–605 (2003).
11. P. Parimi *et al.*, *Nature* **426**, 404 (2003).
12. A. Poddubny, I. Iorsh, P. Belov, Y. Kivshar, *Nat. Photon.* **7**, 948-957 (2013).
13. J. S. G.-D, A. Alù, *ACS Photonics* **3**, 2211–2224 (2016).
14. T. Xu *et al.*, *Nature* **497**, 470–474 (2013).

15. H. J. Lezec, J. A. Dionne, H. A. Atwater, *Science* **316**, 430-432 (2007).
16. H. Shin, S. Fan, *Phys. Rev. Lett.* **96**, 073907 (2006).
17. A. Vakil, N. Engheta, *Science* **332**, 1291-1294 (2011).
18. D. N. Basov, M. M. Fogler, F. J. Garcia de Abajo, *Science* **354**, aag1992 (2016).
19. T. Low *et al.*, *Nat. Mater.* **16**, 182-194 (2017).
20. Q. Zhang *et al.*, *Nature* **597**, 187-195 (2021).
21. Y. Wu *et al.*, *Nat. Rev. Phys.* **4**, 578-594 (2022).
22. A. D. L. K. V. Sreekanth, and G. Strangi, *Appl. Phys. Lett.* **103**, 023107 (2013).
23. X. Lin *et al.*, *Proc. Natl. Acad. Sci.* **114**, 6717-6721 (2017).
24. P. Alonso-González *et al.*, *Science* **344**, 1369-1373 (2014).
25. P. Li *et al.*, *Science* **359**, 892-896 (2018).
26. W. Ma *et al.*, *Nature* **562**, 557-562 (2018).
27. Z. Zheng *et al.*, *Sci. Adv.* **5**, eaav8690 (2019).
28. G. Hu *et al.*, *Nature* **582**, 209-213 (2020).
29. M. Chen *et al.*, *Nat. Mater.* **19**, 1307-1311 (2020).
30. J. Chen et al., *Nature* **487**, 77-81 (2012).
31. Z. Fei *et al.*, *Nature* **487**, 82-85 (2012).
32. G. Álvarez-Pérez *et al.*, *ACS Photonics* **9**, 383-390 (2022).
33. Y. Zeng *et al.*, *Nano Lett.* **22**, 4260-4268 (2022).
34. F. L. Ruta *et al.*, *Nat. Commun.* **13**, 3719 (2022).
35. H. Hu *et al.*, *Nat. Nanotechnol.* **17**, 940–946 (2022).
36. Gonzalo Álvarez-Pérez, *et al.*, *Sci. Adv.* **8**, abp8486 (2022).
37. Y. Zhang *et al.*, *Nature* **459**, 820-823 (2009).
38. D. J. Rizzo *et al.*, *Nano Lett.* **20**, 8438-8445 (2020).

# Supplementary Information

**This PDF file includes:**

**Supplementary Figure 1.** Illustration of the device structure used in our theoretical model.

**Note 1.** Calculation of the dispersion and IFCs of hybrid plasmon-phonon polaritons.

**Supplementary Figure 2.** Theoretical analysis of incidence and refraction angles.

**Note 2.** Theoretical analysis of the focal position associated with negative refraction.

**Supplementary Figure 3.** Illustration of the dry transfer process.

**Supplementary Figure 4.** Illustration of AFM probe transfer method.

**Supplementary Figure 5.** Simulation and theoretical analysis of the relationship between focal position and launching source position.

**Supplementary Figure 6.** Experimentally measured near-field distributions of polaritons and corresponding simulated field distributions $Re\{E_z\}$.

**Supplementary Figure 7.** Simulation of modal-profile matching.

**Supplementary Figure 8.** Analysis of optical losses at the graphene interface.

**Supplementary Figure 9.** All-angle negative refraction over a wide spectral range.

**Supplementary Figure 10.** Experimental demonstration of relationship between focal position and various illumination frequencies.

**Supplementary Figure 11.** Negative refraction by negative group velocity of reversed dispersion.

**Supplementary Figure 12.** Gate-tunable hybrid polaritons.

**Supplementary Figure 13.** Gate-tunable negative refraction with different graphene Fermi energies.

**Supplementary Figure 14.** Extraction analysis of FWHM and intensity enhancement of the focal spots for various gate voltages.

**Supplementary Figure 15.** Gate-tunable negative refraction of another sample.

**Supplementary Figure 16.** Effect of monolayer $\alpha$-RuCl$_3$ on the propagation of hybrid polaritons.

**References**

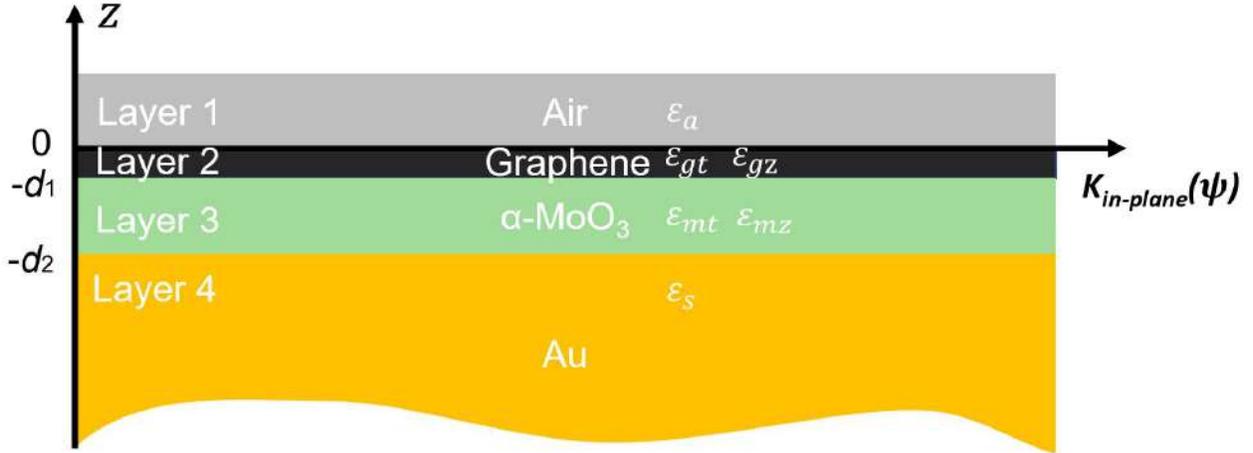

**Supplementary Figure 1. Illustration of the device structure used in our theoretical model.** Layers 1 to 4 correspond to air, graphene, α-MoO₃, and Au, respectively. Each layer is described by its dielectric tensor.

## Note 1. Calculation of the dispersion and IFCs of hybrid plasmon-phonon polaritons.

In the theoretical model, we treat the graphene/α-MoO₃ heterostructure as a laterally-infinite stratified medium consisting of four layers: $z > 0$ (air, layer 1), $-d_1 < z < 0$ (graphene, layer 2), $-d_2 < z < -d_1$ (α-MoO₃, layer 3), and $z < -d_2$ (Au, layer 4). Each layer is normal to the $z$ axis and represented by its corresponding dielectric tensor. The air and Au layers are described by isotropic tensors $\text{diag}\{\varepsilon_{a,s}\}$, while graphene is modeled as a uniaxial anisotropic material (*1*), whose permittivity tensor is given by

$$\bar{\bar{\varepsilon}}_{graphene} = \begin{bmatrix} \varepsilon_{gt} & 0 & 0 \\ 0 & \varepsilon_{gt} & 0 \\ 0 & 0 & \varepsilon_{gz} \end{bmatrix}, \qquad (S1)$$

where the tangential permittivity is expressed as

$$\varepsilon_{gt} = 1 + i \frac{i\sigma}{\omega \varepsilon_0 t_g} \qquad (S2)$$

while the out-of-plane permittivity is set to

$$\varepsilon_{gz} = 1. \qquad (S3)$$

Here, $\varepsilon_0$ is the free-space permittivity, $t_g$ represents the thickness of monolayer graphene, which is set to 0.33 nm (the interlayer spacing in graphite), and $\sigma$ represents the optical conductivity of graphene taken from the local limit of the random-phase approximation (RPA) (*2*).

As α-MoO₃ is a biaxial anisotropic material, its permittivity tensor is modeled as

$$\bar{\bar{\varepsilon}}_{\text{MoO}_3} = \begin{bmatrix} \varepsilon_{mx} & 0 & 0 \\ 0 & \varepsilon_{my} & 0 \\ 0 & 0 & \varepsilon_{mz} \end{bmatrix}, \tag{S4}$$

where $\varepsilon_{mx}$, $\varepsilon_{my}$, and $\varepsilon_{mz}$ are the permittivity components along the principal axes (3).

Without loss of generality, we consider polariton waves propagating with an in-plane wave vector $\vec{k}_{in-plane} = (k_x, k_y)$, such that

$$k_x = k_{\text{in-plane}} \cos(\psi), \qquad k_y = k_{\text{in-plane}} \sin(\psi), \tag{S5}$$

where $\psi$ denotes the angle between $\vec{k}_{in-plane}$ and the x axis.

Through a simple coordinate transformation, the dielectric function of α-MoO₃ along the direction of wave propagation $\vec{k}_{\text{in-plane}}$ can be expressed as $\varepsilon_{mt} = \varepsilon_{mx} \cos^2(\psi) + \varepsilon_{my} \sin^2(\psi)$, and thus, we find a 2D dielectric function diag $\{\varepsilon_{mt}, \varepsilon_{mz}\}$. Similarly, we can write the graphene permittivity tensor as diag $\{\varepsilon_{gt}, \varepsilon_{gz}\}$ (see Supplementary Figure 1).

A typical polariton waveguide mode can have TM, TE, or mixed TM-TE symmetry. In our system, the α-MoO₃ layer contains TM-TE mixed modes, but the weight of the TE component is much weaker than the TM one (4). As an approximation, for the sake of simplicity, we consider TM modes in the derivation that follows. We focus on a 2D waveguide mode with $\vec{k}_{\text{in-plane}}$ directed along the x axis, as shown in Supplementary Figure 1. The transverse magnetic field in the graphene/α-MoO₃ heterostructure can be expressed as $\vec{H}_y = H_y(z) \exp(iqx - \omega t) \, \vec{e}_y$, where $q = |\vec{k}_{in-plane}|$ and $\vec{e}_y$ is the unit vector along y. Solving the wave equation $\nabla^2 \vec{H} + k_0^2 \, \bar{\bar{\varepsilon}} \vec{H} = \nabla(\nabla \cdot \vec{H})$ in each layer, we find the relations

$$\partial^2 \vec{H}_y + \left( k_0^2 \varepsilon_{a,s} - q^2 \right) \vec{H}_y = 0, \tag{S6}$$

$$\partial^2 \vec{H}_y + \left( k_0^2 \varepsilon_t^{(j)} - \left( \frac{\varepsilon_t^{(j)}}{\varepsilon_z^{(j)}} \right) q^2 \right) \vec{H}_y = 0, \tag{S7}$$

where j = 2 and j = 3 denote the graphene and α-MoO₃ layers, respectively. For a van der Waals structure of finite thickness, the solution for $H_y(z)$ can be expressed as

$$H_y(z) = \begin{cases} (A+B)e^{-\alpha_a z} & (z > 0), \\ Ae^{ik_z^{(2)}z} + Be^{-ik_z^{(2)}z} & (-d_1 < z \leq 0), \\ Ce^{ik_z^{(3)}z} + De^{-ik_z^{(3)}z} & (-d_2 < z \leq -d_1), \\ \left(Ce^{-ik_z^{(3)}d_2} + De^{ik_z^{(3)}d_2}\right)e^{\alpha_s(z+d_2)} & (z \leq -d_2). \end{cases} \quad (S8)$$

Here, $\alpha_{a,s} = \sqrt{q^2 - k_0^2 \varepsilon_{a,s}}$ is the out-of-plane decay coefficient of the air layer ($j = 1$, $\alpha_a$) and Au layer ($j = 4$, $\alpha_s$), while $k_z^{(j)} = \sqrt{k_0^2 \varepsilon_t^{(j)} - \left(\frac{\varepsilon_t^{(j)}}{\varepsilon_z^{(j)}}\right)q^2}$ is the normal wave vector component in the graphene ($j = 2$) and α-MoO₃ ($j = 3$) layers. The four unknown parameters $A$, $B$, $C$, and $D$ are the amplitude coefficients of the nontrivial solution that is found by matching the electromagnetic boundary conditions and imposing the vanishing of the associated secular determinant. In addition, $d_1$ and $d_2$ are the thicknesses of the graphene ($j = 2$) and α-MoO₃ ($j = 3$) layers, respectively.

To find $\vec{E}_x = E_x(z)\exp(iqx - \omega t)\vec{e}_x$, we plug $H_y$ into the curl equation $\nabla \times H = i\omega\varepsilon_0\bar{\bar{\varepsilon}}E$. We obtain

$$E_x(z) = \begin{cases} \dfrac{i\alpha_a}{\omega\varepsilon_0\varepsilon_a}(A+B)e^{-\alpha_a z} & (z > 0), \\ \dfrac{1}{\omega\varepsilon_0\varepsilon_t^{(2)}}\left(Ak_z^{(2)}e^{ik_z^{(2)}z} - Bk_z^{(2)}e^{-ik_z^{(2)}z}\right) & (-d_1 < z \leq 0), \\ \dfrac{1}{\omega\varepsilon_0\varepsilon_t^{(3)}}\left(Ck_z^{(3)}e^{ik_z^{(3)}z} - Dk_z^{(3)}e^{-ik_z^{(3)}z}\right) & (-d_2 < z \leq -d_1), \\ \dfrac{-i\alpha_s}{\omega\varepsilon_0\varepsilon_s}\left(Ce^{-ik_z^{(3)}d_2} + De^{ik_z^{(3)}d_2}\right)e^{\alpha_s(z+d_2)} & (z \leq -d_2). \end{cases} \quad (S9)$$

Considering that the tangential components of $\vec{E}$ and $\vec{H}$ are continuous at the interfaces, we find the equations

$$\begin{cases} E_x^{(1)} = E_x^{(2)}, H_y^{(1)} = H_y^{(2)}, & z_1 = 0, \\ E_x^{(2)} = E_x^{(3)}, H_y^{(2)} = H_y^{(3)}, & z_2 = -d_1, \\ E_x^{(3)} = E_x^{(4)}, H_y^{(3)} = H_y^{(4)}, & z_3 = -d_2. \end{cases} \quad (S10)$$

By substituting the fields in Eqs. (S8)-(S9) into the boundary conditions, we obtain the secular matrix equation

$$\begin{bmatrix} \dfrac{i\alpha_a}{\varepsilon_a} - \dfrac{k_z^{(2)}}{\varepsilon_t^{(2)}} & \dfrac{i\alpha_a}{\varepsilon_a} + \dfrac{k_z^{(2)}}{\varepsilon_t^{(2)}} & 0 & 0 \\[2mm] \dfrac{k_z^{(2)}}{\varepsilon_t^{(2)}} e^{-ik_z^{(2)}d_1} & -\dfrac{k_z^{(2)}}{\varepsilon_t^{(2)}} e^{ik_z^{(2)}d_1} & -\dfrac{k_z^{(3)}}{\varepsilon_t^{(3)}} e^{-ik_z^{(3)}d_1} & \dfrac{k_z^{(3)}}{\varepsilon_t^{(3)}} e^{ik_z^{(3)}d_1} \\[2mm] e^{-ik_z^{(2)}d_1} & e^{ik_z^{(2)}d_1} & -e^{-ik_z^{(3)}d_1} & e^{ik_z^{(3)}d_1} \\[2mm] 0 & 0 & \left(\dfrac{i\alpha_s}{\varepsilon_s} + \dfrac{k_z^{(3)}}{\varepsilon_t^{(3)}}\right) e^{-ik_z^{(3)}d_2} & \left(\dfrac{i\alpha_s}{\varepsilon_s} - \dfrac{k_z^{(3)}}{\varepsilon_t^{(3)}}\right) e^{ik_z^{(3)}d_2} \end{bmatrix} \begin{bmatrix} A \\ B \\ C \\ D \end{bmatrix} = 0. \quad (S11)$$

Denoting the square matrix in this equation by M, we impose the vanishing of the determinant det{M} = 0 to guarantee the existence of a nonzero solution for the amplitude coefficients $A$, $B$, $C$, and $D$. Such condition leads to the transcendental equation

$$e^{2p_1} = -\frac{[(p_2-p_3)(p_3-p_4)(p_4+p_5)]e^{ik_z^{(2)}d_1} + [(p_2+p_3)(p_3+p_4)(p_4+p_5)]e^{-ik_z^{(2)}d_1}}{[(p_2-p_3)(p_3+p_4)(p_4-p_5)]e^{ik_z^{(2)}d_1} + [(p_2+p_3)(p_3-p_4)(p_4-p_5)]e^{-ik_z^{(2)}d_1}}, \quad (S12)$$

where

$$\begin{aligned} p_1 &= -ik_z^{(3)}d_1 + ik_z^{(3)}d_2, \\ p_2 &= \frac{i\alpha_a}{\varepsilon_a}, \\ p_3 &= \frac{k_z^{(2)}}{\varepsilon_t^{(2)}}, \\ p_4 &= \frac{k_z^{(3)}}{\varepsilon_t^{(3)}}, \\ p_5 &= \frac{i\alpha_s}{\varepsilon_s}. \end{aligned} \quad (S13)$$

The solution to Eq. (S12) yields the mode dispersion relation, which can be solved for each value of the angle $\psi$ (see Eq. (S5)), thus generating a function $q(\psi)$ that can be understood as for isofrequency contour (IFC) of the polaritons supported by the graphene/α-MoO₃ heterostructure. Following this procedure, we find the numerical results that are shown in Figure 1B.

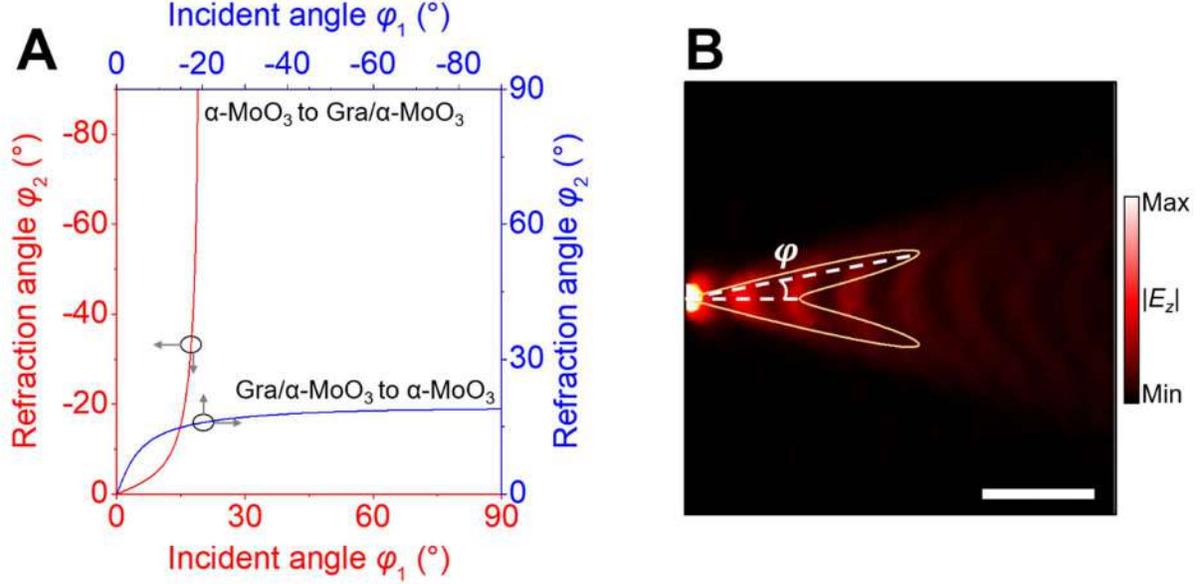

**Supplementary Figure 2. Theoretical analysis of the incidence and refraction angles. (A)** Incidence angle $\varphi_1$ as a function of refraction angle $\varphi_2$ for negative refraction after transmission through the in-plane interface for both directions of propagation (see labels). These angles refer to the incidence and refracted polariton Poynting vector **S**. **(B)** Numerical simulation of the electric field $|E_z|$ (color plot) excited by a point electric dipole placed 100 nm above the α-MoO$_3$ film, together with the polar distribution of the electric field in real space (orange solid curve). $\varphi$ indicates the incidence angle of the polariton Poynting vector **S**. The scale bar indicates 1 μm.

## Note 2. Theoretical analysis of the focal position associated with negative refraction.

As the tangential wave vectors of the interface for the refraction process are continuous (this is essentially Snell's Law), the relationship between the incidence and refraction angles can be obtained numerically, as illustrated in Supplementary Figure 2A.

For the bare α-MoO$_3$ film, the hyperbolic polaritons in the Reststrahlen band II (816~976 cm$^{-1}$) are highly oriented. When approaching the two asymptotes of the hyperbolic IFC, the number of available wave vectors of PhPs and the intensity of the electric field are significantly increased, as inferred upon examination of the Dyadic Green tensor (5). More precisely, the $zz$ component of the Dyadic Green tensor of the bare α-MoO$_3$ film reduces to

$$G_{zz}(\mathbf{r} - \mathbf{r}') = \frac{\epsilon^{\frac{3}{2}}}{\sqrt{(\pi d)}} \frac{(\epsilon_x^2 \Delta y^2 + \epsilon_y^2 \Delta x^2) e^{-qk_0 H}}{\epsilon_x^2 \epsilon_y^{\frac{5}{4}} (k_0 d)^2 (\epsilon_x \Delta y^2 + \epsilon_y \Delta x^2)^{\frac{5}{4}}} \exp\left\{ \frac{i\left(2\epsilon\sqrt{\epsilon_x \Delta y^2 + \epsilon_y \Delta x^2}\right)}{\epsilon_x \sqrt{\epsilon_y} d} - \frac{\pi}{4} \right\}, \text{(S14)}$$

where

$$q = -\frac{2\epsilon\sqrt{\epsilon_x^2\Delta y^2 + \epsilon_y^2\Delta x^2}}{\epsilon_x\sqrt{\epsilon_y}k_0 d\sqrt{\epsilon_x\Delta y^2 + \epsilon_y\Delta x^2}} \qquad (S15)$$

denotes the in-plane wave vector and $\epsilon$ represents the relative dielectric function given by the average value of the upper and lower media. We take $\epsilon = 1$ and denote $\Delta r = |\mathbf{r} - \mathbf{r'}|$ the distance from the target point to the source point, with relative vector projections on the $x$ and $y$ directions given by $\Delta x$ and $\Delta y$. In addition, $k_0$ represents the free-space wave vector of the incident light, $d$ denotes the thickness of α-MoO₃, and $H$ is the height of the detection point.

In Supplementary Figure 2B, we plot the absolute value of $G_{zz}(r - r')$ (orange solid curve), obtained from Eq. (S14) as a function of polar angle at a distance $\Delta r = 2.5$ μm from the source point for an illumination frequency of 893 cm⁻¹. The thickness of α-MoO₃ is set at $t = 240$ nm. We set the monitoring point for the α-MoO₃ film at a height $H = 50$ nm. The pseudo-color map of the background is obtained from a numerical simulation of the electric-field absolute amplitude $|E_z|$ excited by a point electric dipole placed 100 nm above the α-MoO₃ surface. The near-field absolute amplitude shows a maximum at an angle $\varphi_c$ with respect to the [100] crystal direction (set to $x$ here), which indicates that energy is mainly canalized along this direction. We consider $\varphi_c$ as the main angle sustained by the incident light, such that the intersection of the refraction angle corresponding to $\varphi_f$ is calculated as the focal point. Using this simplified model, the relationship between the positions of the source and the focal spot can be determined to be

$$d = f\,\frac{\tan(\varphi_c)}{\tan(\varphi_f)}, \qquad (S16)$$

where $d$ and $f$ represent the distance from the source point and the focal spot to the graphene edge, respectively.

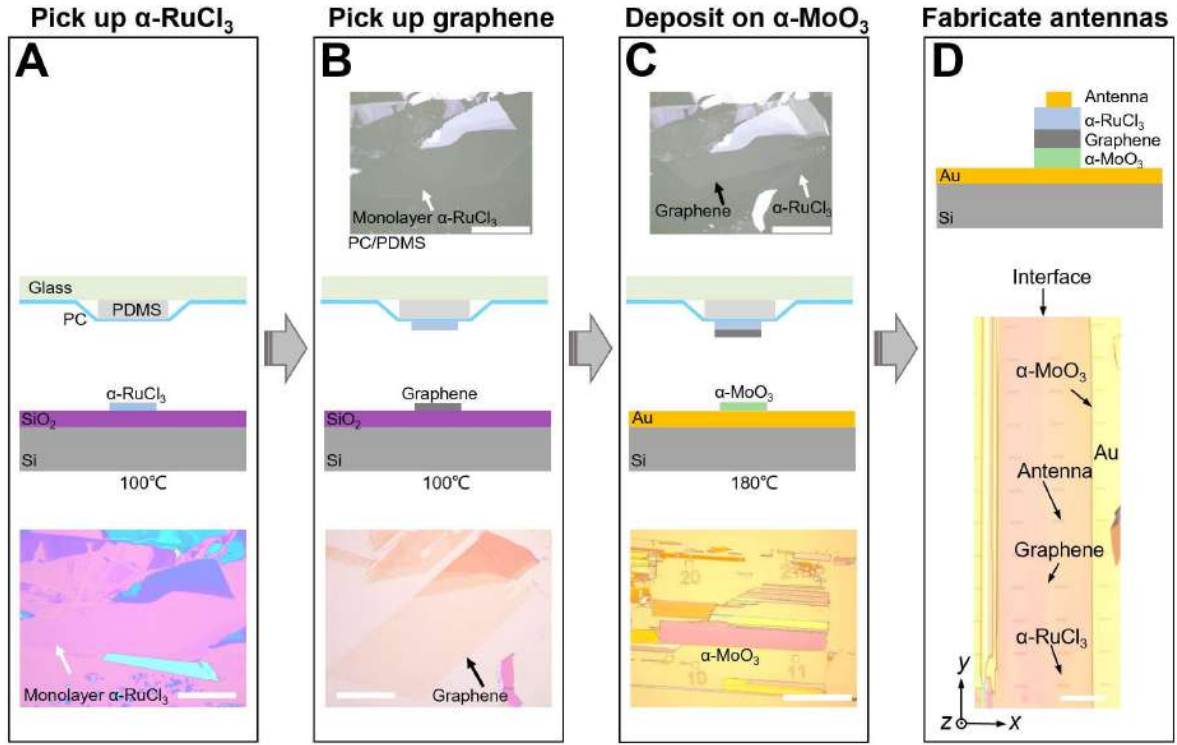

**Supplementary Figure 3. Illustration of the dry transfer process. (A)** A glass/PDMS/PC stamp is brought into contact with a monolayer of α-RuCl₃ exfoliated on a SiO₂ (300 nm)/Si substrate at 100°C and then withdrawn. **(B)** The stamp along with α-RuCl₃ is contracted with an exfoliated graphene monolayer at 100 °C, and again slowly withdrawn. **(C, D)** The α-RuCl₃/graphene is aligned and contracted with an α-MoO₃ film previously exfoliated on a Au (60 nm)/Si substrate. Then, the temperature is raised to 180 °C and the stack is released onto the α-MoO₃ film, forming an α-RuCl₃/graphene/α-MoO₃ heterostructure. Scale bars indicate 20 µm in panels (A-C) and 8 µm in panel (D).

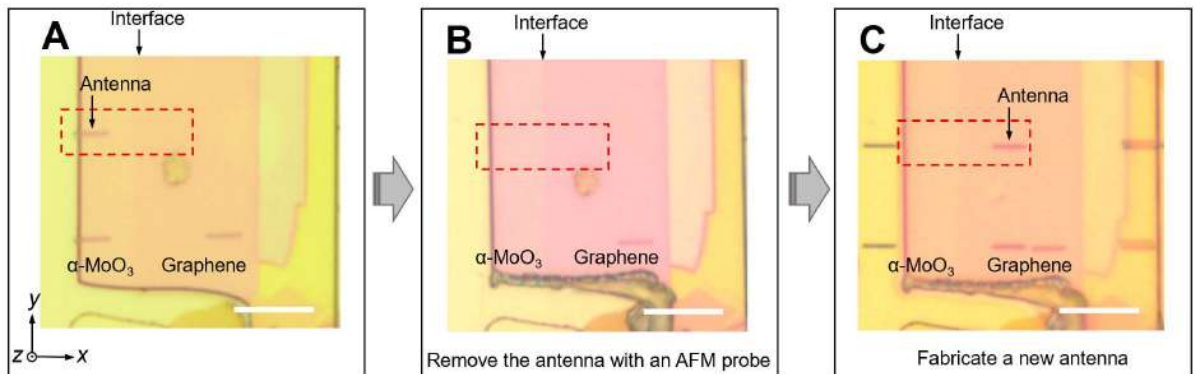

**Supplementary Figure 4. Illustration of AFM probe transfer method. (A)** Gold antenna arrays on the α-MoO₃ film. **(B)** The gold antennas on α-MoO₃ are removed away by an AFM tip. **(C)** Another gold antenna array is fabricated on the graphene/α-MoO₃ side. The α-MoO₃ thickness is 242 nm. This sample is used for near-field measurements in Figure 2 in the main text. Scale bars indicate 8 µm.

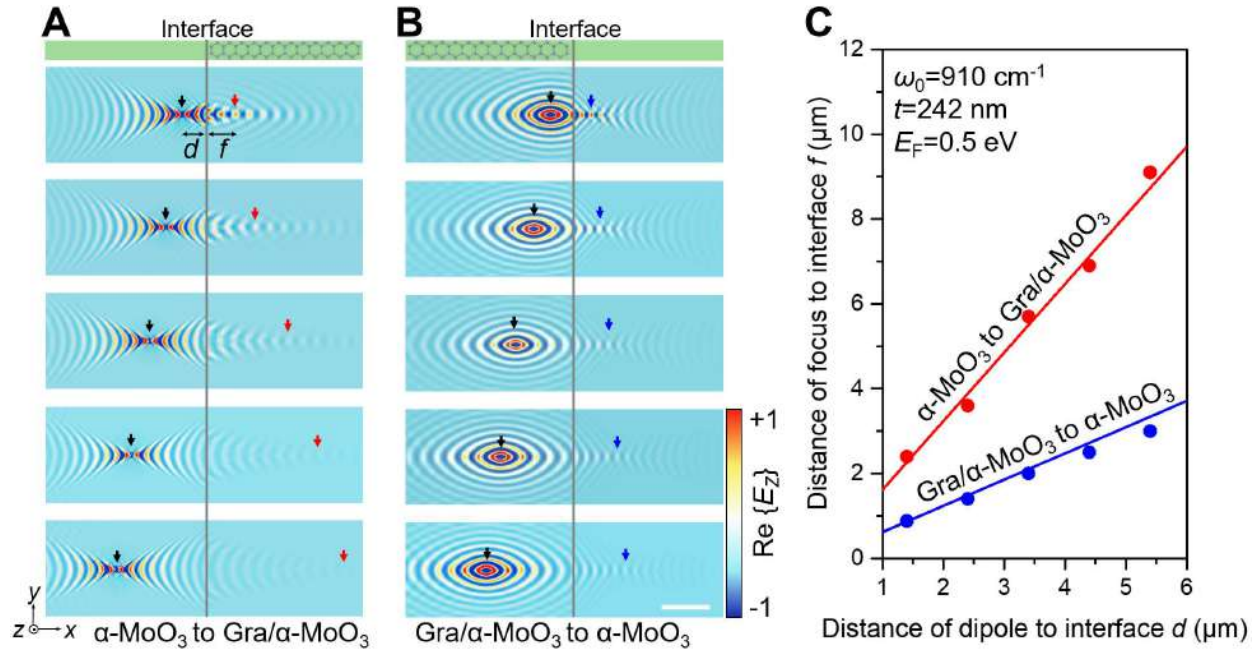

**Supplementary Figure 5. Simulation and theoretical analysis of the relationship between focal position and launching source position. (A)** Simulated spatial distribution of the electric field Re{$E_z$} showing negative refraction from α-MoO₃ with a hyperbolic wave to graphene/α-MoO₃ with an elliptic wave. Here, *d* and *f* represent the distance of the interface to the launching source and focal spot positions (interface-focus distance), respectively. **(B)** Reversible negative refraction relative to panel (A), where the antenna is moved to the graphene/α-MoO₃ side. The black arrows indicate the position of the emitting dipole. The red and blue arrows mark the focal spot. The scale bar indicates 2 μm. **(C)** *d* as a function of *f* for the configurations in panels (A) (red) and (B) (blue). The dots are extracted from the simulations presented in panels (A, B), while the solid lines stand for the theoretical analysis based on the method described in Note 2.

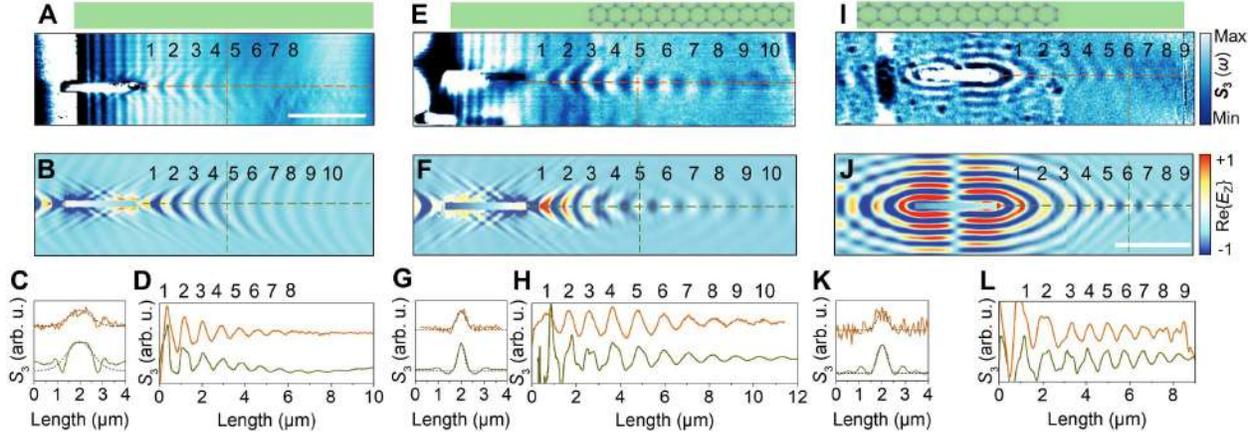

**Supplementary Figure 6. Experimentally measured near-field distributions of polaritons and corresponding simulated field distributions Re{$E_z$}. (A-D)** Hyperbolic polaritons launched and propagated in bare α-MoO₃ as a control experiment. **(E-H)** Hyperbolic polaritons launched from α-MoO₃ to the interface with the graphene edge, undergoing negative refraction and focusing. **(I-L)** Reversed negative refraction and focusing relative to panels (E-H), where the antenna is placed on the graphene/α-MoO₃ side to launch hybrid plasmon-phonon polaritons. **(A, E, I)** Experimentally measured near-field amplitude of polaritons. **(B, F, J)** Simulated field amplitude Re{$E_z$} corresponding to panels (A, E, I). **(C, G, K)** Near-field profiles of the signal along the orange and green vertical dashed lines in panels (A, B), (E, F), and (I, J), respectively. The black dashed curves are fitting Gaussians. **(D, H, L)** Near-field profiles of the signal along the orange and green horizontal dashed lines in panels (A, B), (E, F), and (I, J), respectively. Horizontal green stripes on top of the images represent the α-MoO₃ film. All the experimental results are measured from the *in-situ* sample shown in Supplementary Figure 4. The thickness of α-MoO₃ is 242 nm. The scale bar indicates 3 μm. The graphene is prepared by mechanical exfoliation and statically doped to a Fermi energy $E_F$ = 0.5 eV by adding a monolayer of α-RuCl₃. The illumination frequency is fixed at $\omega_0$ = 893 cm⁻¹.

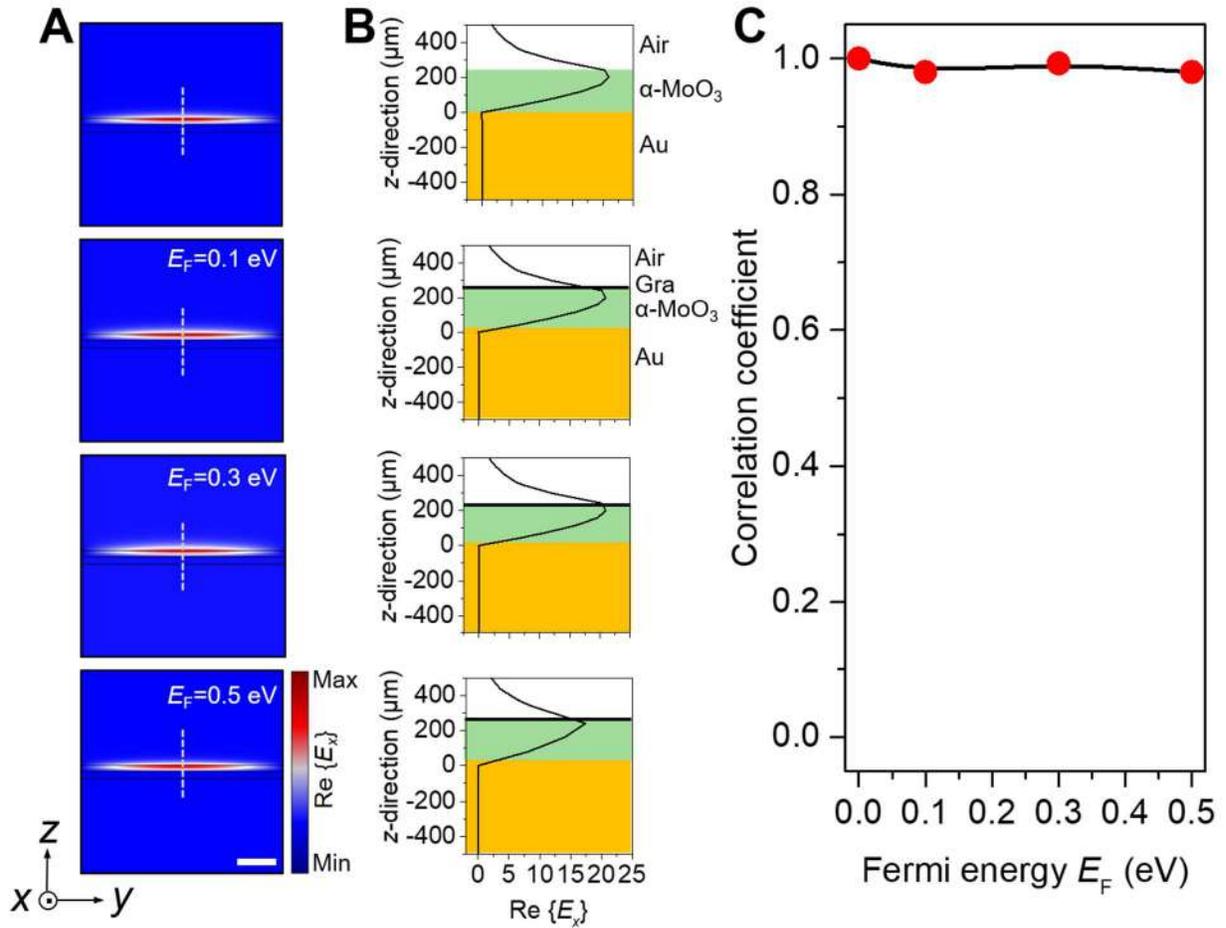

**Supplementary Figure 7. Simulation of modal-profile matching. (A)** Mode profiles for different graphene Fermi energies. The thickness of α-MoO₃ is 242 nm. The scale bar indicates 3 μm. The illumination frequency is fixed at $\omega_0$=900 cm⁻¹. **(B)** Electric field profiles for different graphene Fermi energies corresponding to panel (A). **(C)** Correlation coefficient (also known as the Pearson correlation coefficient) of the electric field profiles for different graphene Fermi energies compared with that of bare α-MoO₃, shown at the top of panel (B). The modal-profile mismatch is less than 3% at different graphene Fermi energies due to the strong overlap of the surface modes on both sides.

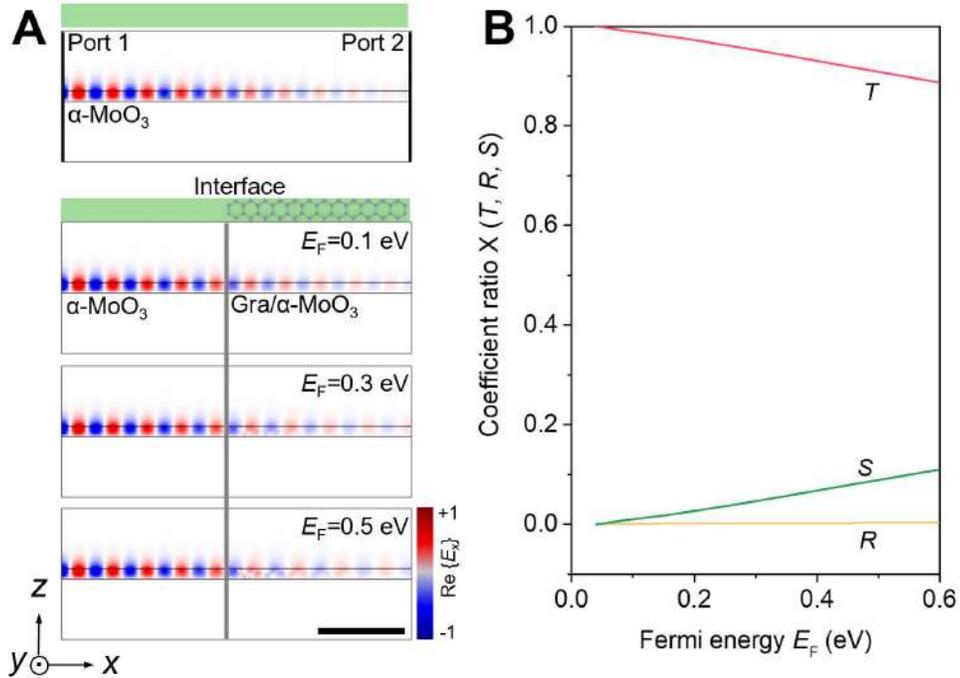

**Supplementary Figure 8. Analysis of optical losses at the graphene interface. (A)** Simulated spatial distribution of the electric field along the $x$ direction as polaritons propagate from α-MoO$_3$ to graphene/α-MoO$_3$ for different graphene Fermi energies (see labels). The thickness of α-MoO$_3$ is 242 nm. The scale bar indicates 2 μm. The illumination frequency is fixed at $\omega_0$ = 900 cm$^{-1}$. **(B)** Polaritonic reflectance (yellow), transmittance (red), and scattering (green) as a function of graphene Fermi energy. Thanks to the single-atom thickness of graphene, losses stemming from scattering at the interface are expected to be negligible and, although they increase with the graphene Fermi energy, the overall loss does not exceed 12%.

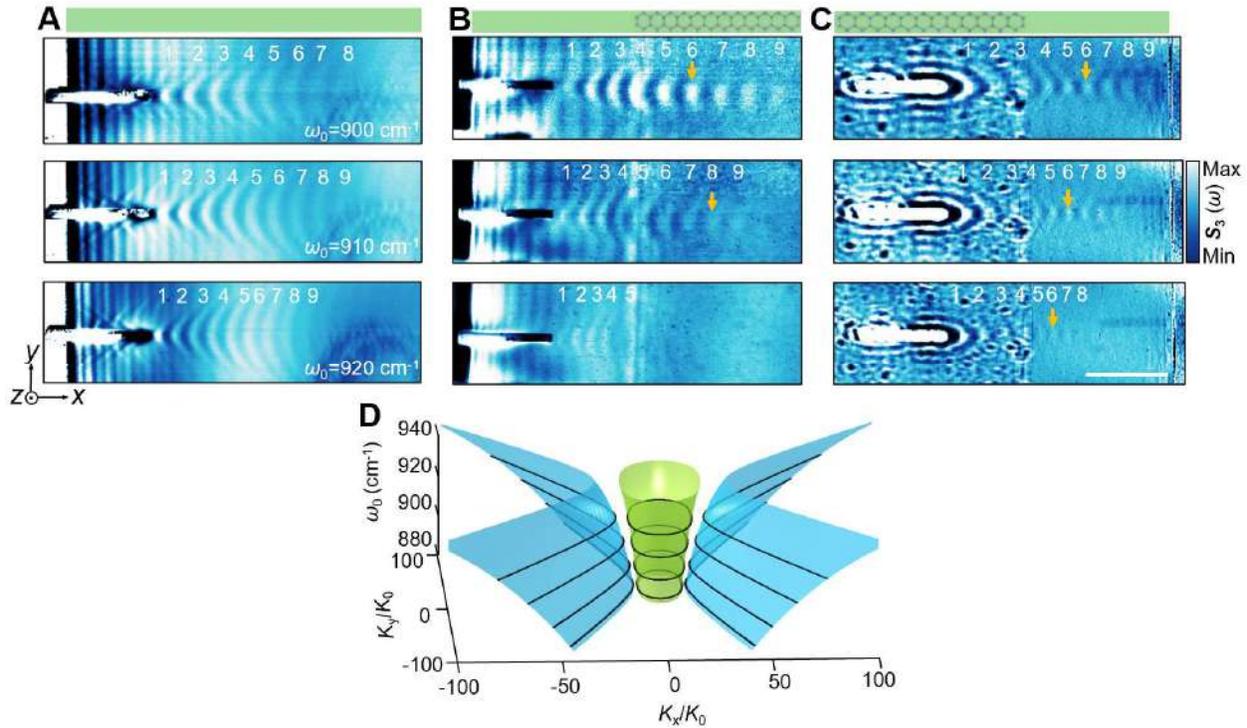

**Supplementary Figure 9. All-angle negative refraction over a wide spectral range.**
**(A)** Hyperbolic polaritons launched and propagated in natural α-MoO$_3$ as a control experiment. (**B**) Measured near-field images showing negative refraction from α-MoO$_3$ to graphene/α-MoO$_3$ with illumination frequencies ranging from 900 to 920 cm$^{-1}$. (**C**) Negative refraction from graphene/α-MoO$_3$ to α-MoO$_3$ in horizontally flipped samples with respect to (B). A launching gold antenna is placed ~3 μm to the left of the interface in all cases. All experimental results are measured from the same *in-situ* sample as in Figure 2. The scale bar indicates 3 μm. (**D**) Three-dimensional representation of the polariton IFCs in α-MoO$_3$ (blue) and graphene/α-MoO$_3$ heterostructure (green), with experimentally measured frequencies represented by black curves. The opening angle *β* of PhPs in α-MoO$_3$ increases with the illumination frequency, which weakens the focusing effect produced by negative refraction from α-MoO$_3$ to graphene/α-MoO$_3$ because of the increased spatial incidence range of PhPs in α-MoO$_3$. In contrast, focusing should be enhanced with an increase of frequency for negative refraction from graphene/α-MoO$_3$ to α-MoO$_3$ due to the increase in refraction angle, which is related to the opening angle.

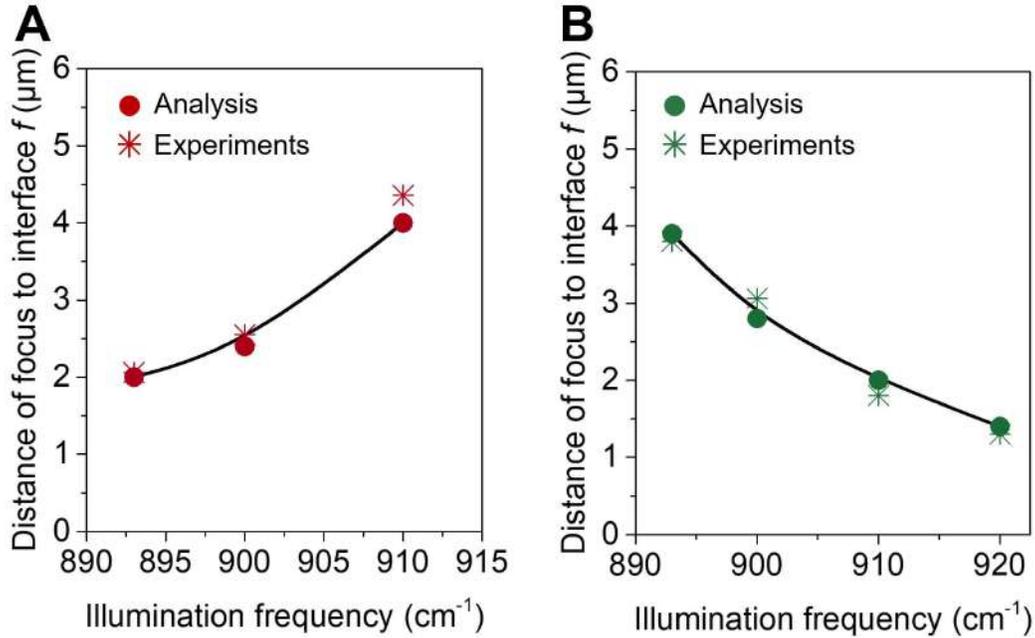

**Supplementary Figure 10. Experimental demonstration of relationship between focal position and various illumination frequencies. (A, B)** Focal distance from the interface as a function of illumination frequency. The red and green dots are extracted from experimental results in Figure 2 and Supplementary Figure 9, while crosses represent theoretical results obtained by using the method described in Note 2, and the solid curves are guides to the eye.

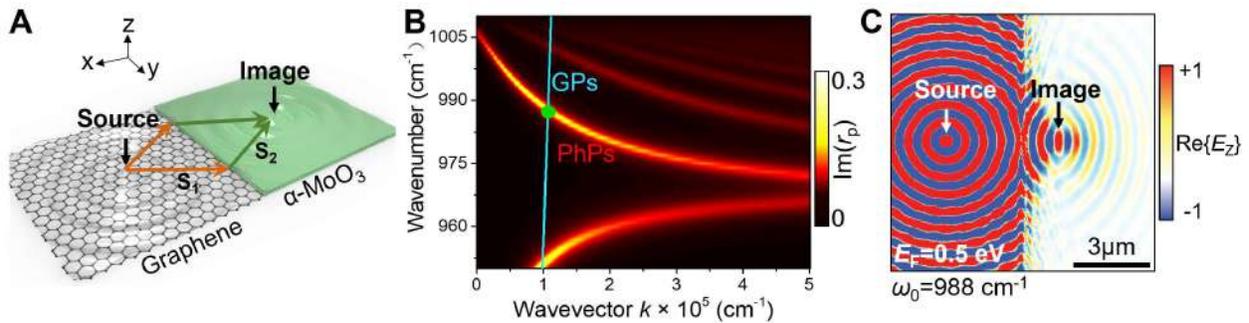

**Supplementary Figure 11. Negative refraction by negative group velocity of reversed dispersion.** (**A**) Schematic of negative refraction between plasmons in graphene and phonon polaritons in α-MoO$_3$ using an in-plane heterostructure. **S**$_1$ and **S**$_2$ indicate the Poynting vectors of plasmon and phonon polaritons, respectively. (**B**) Dispersion of graphene plasmons and phonon polaritons in α-MoO$_3$, corresponding to the left and right region in the panel (A), respectively. (**C**) Simulated spatial distribution of the electric field Re $\{E_z\}$ launched by a dipole source at a frequency of 988 cm$^{-1}$ (corresponding to the green dot in panel (B)), showing negative refraction. The Fermi energy of graphene is set to 0.5 eV and the thickness of α-MoO$_3$ is 260 nm.

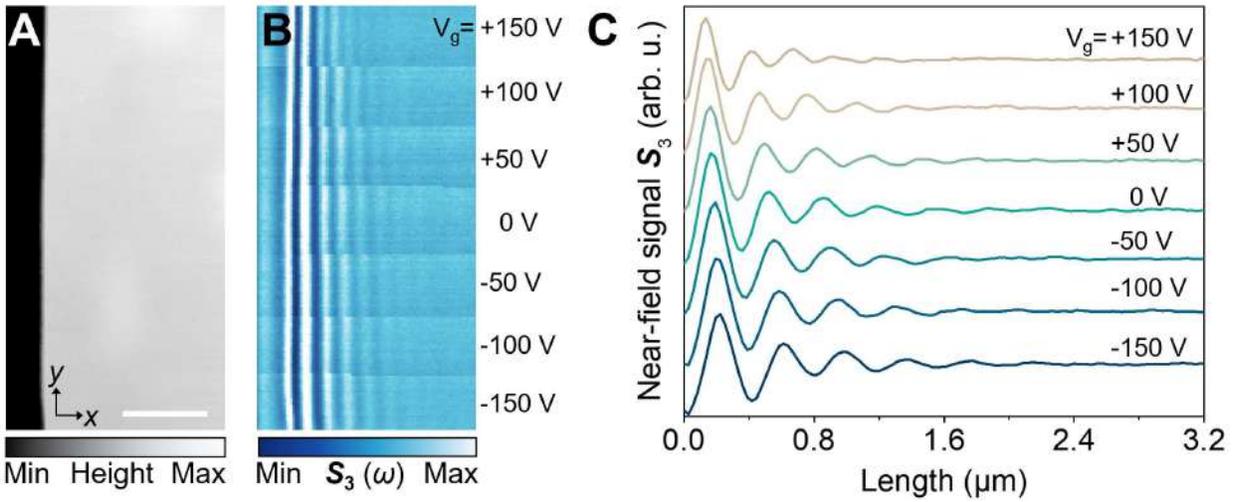

**Supplementary Figure 12. Gate-tunable hybrid polaritons.** (**A**) Topography images of the α-MoO₃ edge corresponding to the samples in Figure 3. (**B**) Real-space infrared nanoimages of gate-tunable hybrid polaritons. (**C**) Near-field profiles extracted from panel (B). The α-MoO₃ thickness is 60 nm and the illumination frequency is 893 $cm^{-1}$. The scale bar indicates 1.5 µm.

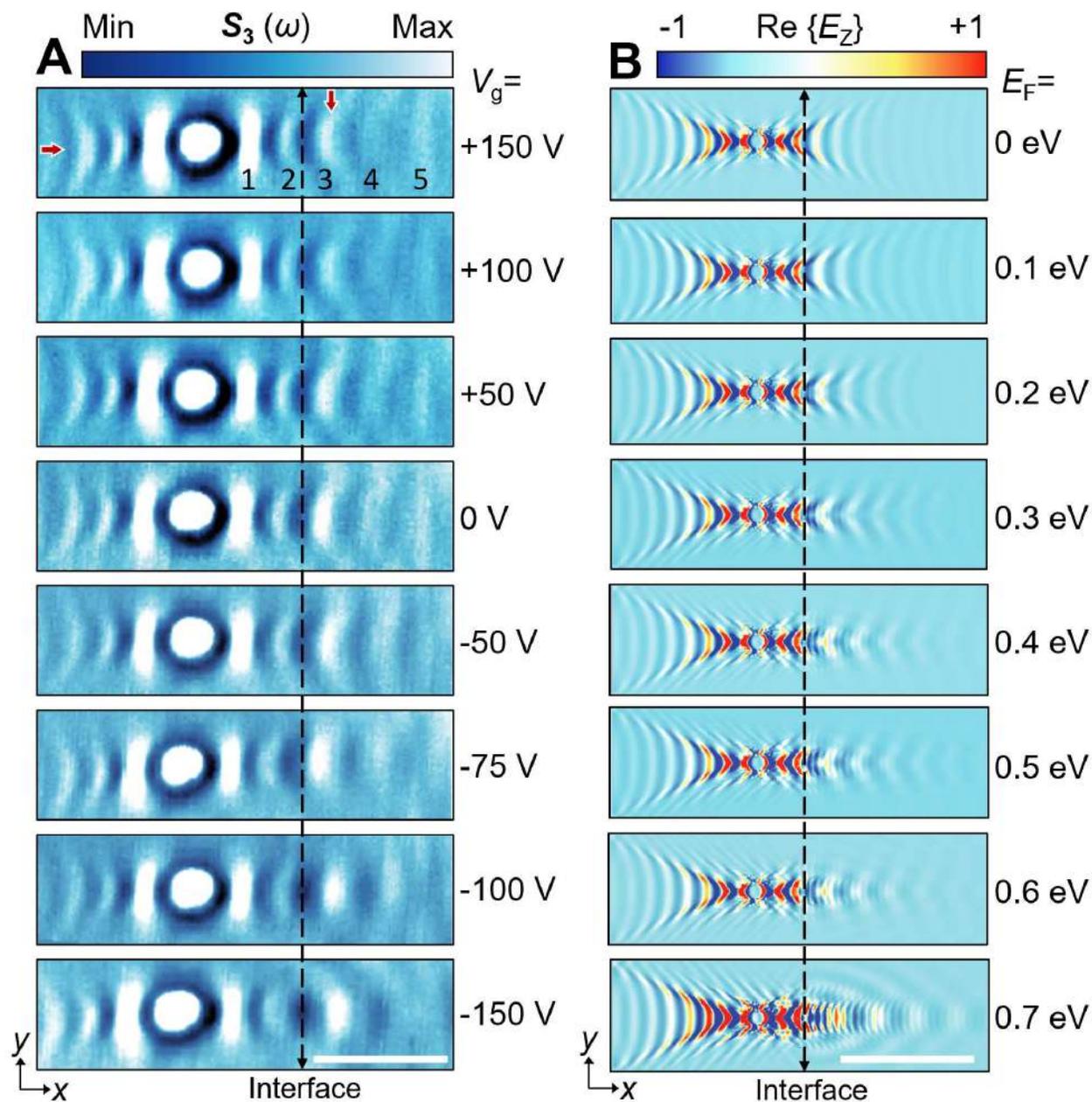

**Supplementary Figure 13. Gate-tunable negative refraction with different graphene Fermi energies. (A)** Experimental near-field images showing gate tunning of negative refraction from α-MoO₃ to graphene/α-MoO₃ with voltages arranged from +150 to -150 V. The scale bar indicates 1.5 μm. **(B)** Numerically simulated negative refraction with various Fermi energies of graphene from $E_F$=0 to 0.7 eV. The scale bar indicates 3 μm. The α-MoO₃ thickness is 60 nm and the illumination frequency is 893 cm⁻¹ in all panels.

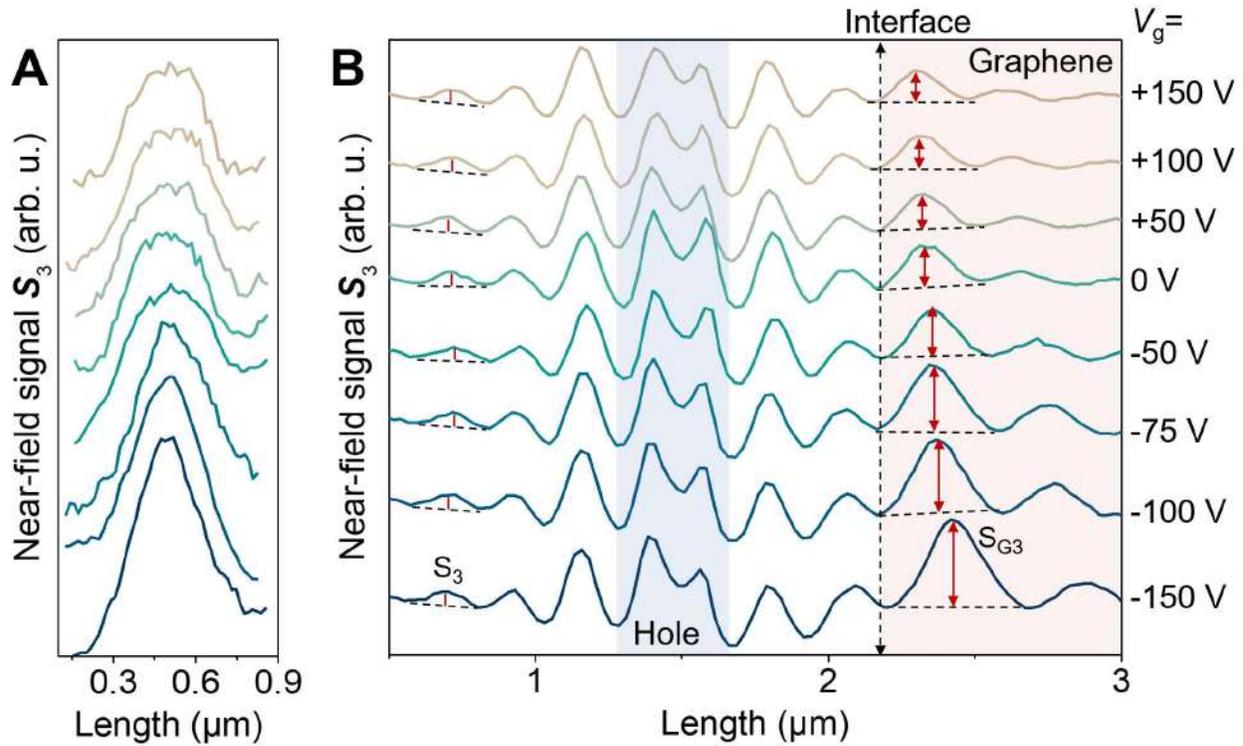

**Supplementary Figure 14. Extraction analysis of FWHM and intensity enhancement of the focal spots for various gate voltages.** FWHM and intensity enhancement are extracted along the *y* and *x* directions (vertical (the third fringe) and horizontal red arrows in Supplementary Figure 13, respectively). The FWHM can be obtained by fitting the near-field signal profiles in panel (A). The intensity enhancement can be determined by comparing the intensity ratio of the focusing fringe to the same order fringe without a negative refraction effect.

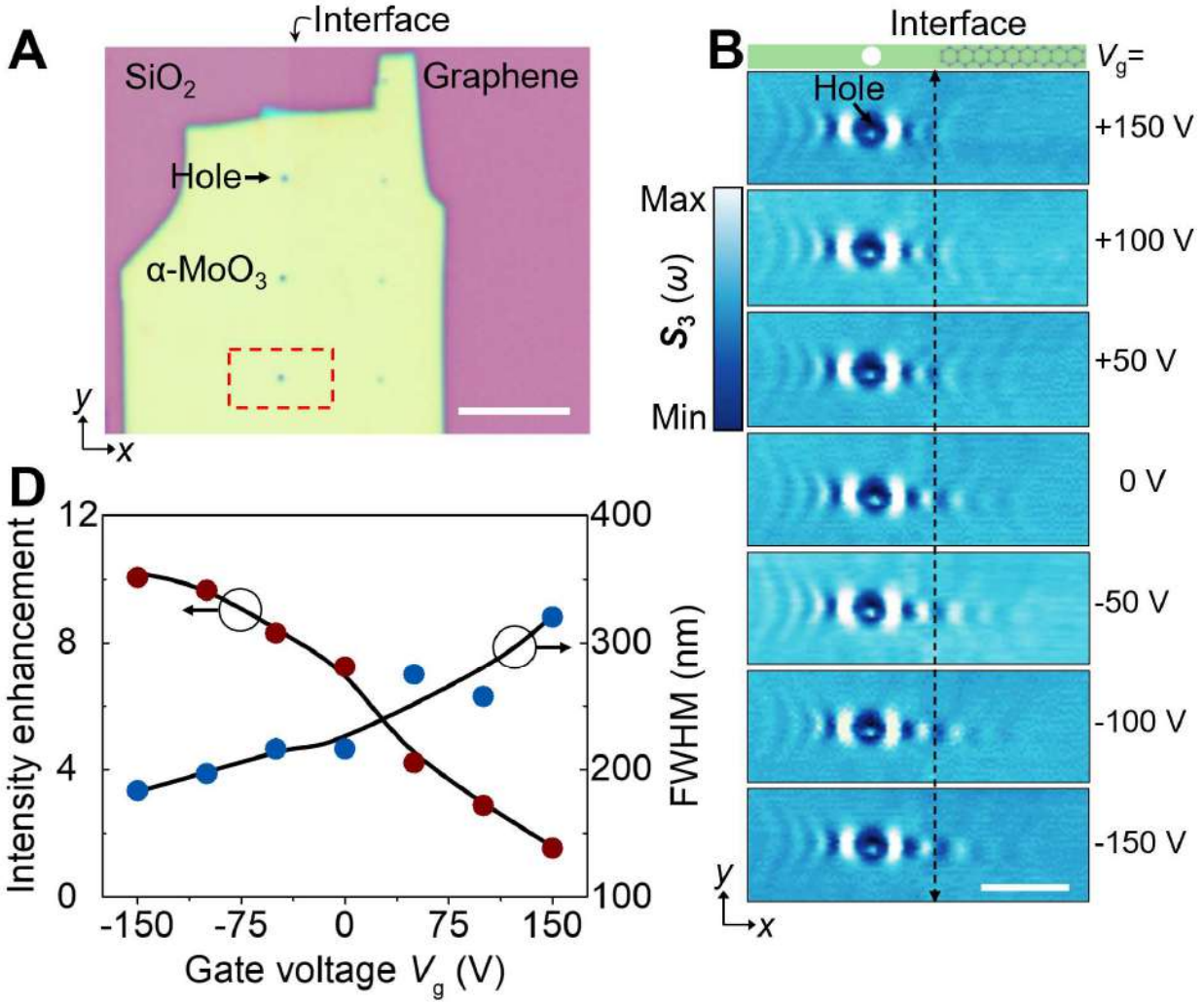

**Supplementary Figure 15. Gate-tunable negative refraction of another sample.** (**A**) Optical image of a gate-tunable device consisting of (from top to bottom) an α-MoO₃ film, monolayer graphene, and SiO₂ substrate. The black horizontal arrow indicates the position of the fabricated circular holes with a diameter of 400 nm. The scale bar stands for 10 μm. (**B**) Experimentally measured near-field images of gate-tunable negative refraction from hyperbolic α-MoO₃ to elliptic graphene/α-MoO₃ with bias voltages varying from +150 V to -150 V. The vertical black dashed line represents the graphene edge. The illumination wavelength is fixed at $\lambda_0$ = 11.20 μm (frequency of 893 $\mathrm{cm}^{-1}$). The scale bar indicates 2 μm. (**C**) Intensity enhancement and FWHM of the focal spots for various gate voltages, taken from the experimental measurements in panel (B).

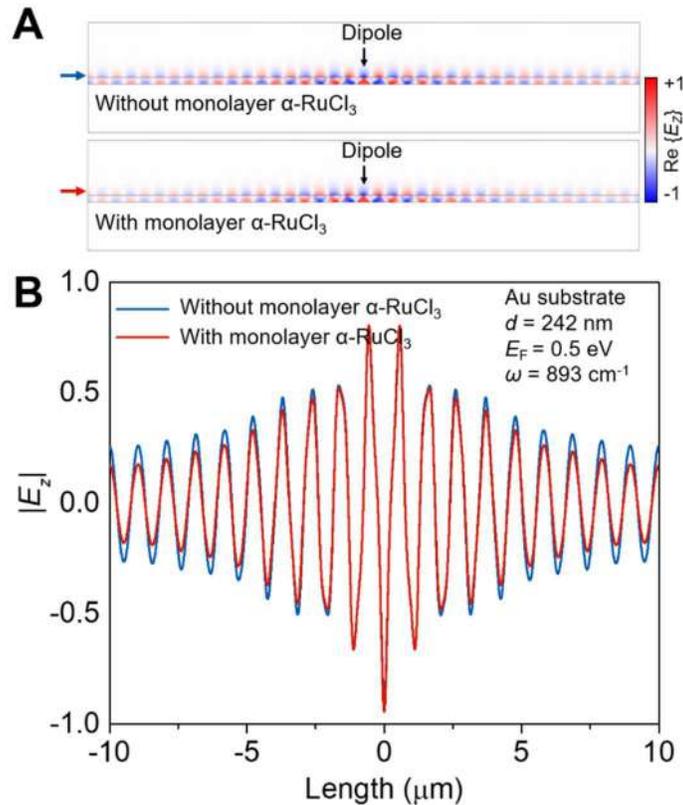

**Supplementary Figure 16. Effect of monolayer α-RuCl₃ on the propagation of hybrid polaritons. (A)** Numerically simulated field distribution of hybrid plasmon-phonon polaritons in graphene/α-MoO₃ with and without monolayer α-RuCl₃. **(B)** Near-field profiles taken at the positions marked by blue and red horizontal arrows in panel (A). The value of $|E_z|$ is normalized to the maximum electric-field intensity. The Fermi energy of graphene is set to $E_F$ = 0.5 eV. The α-MoO₃ thickness is 242 nm. The illumination frequency is 893 cm⁻¹.

**References**

1. Y. Qu *et al*., *Adv. Mater*. **34**, 2105590 (2022).

2. F. J. García de Abajo, *ACS Photonics.* **1**, 135-152 (2014).

3. Z. Zheng *et al*., *Sci. Adv.* **5**, eaav8690 (2019).

4. M. Chen *et al*., *Nat. Mater*. **19**, 1307-1311 (2020).

5. Martín-Sánchez. J *et al*., *Sci. Adv.* **7**, eabj0127 (2021).